# The formation of disk galaxies in a cosmological context

## Populations, metallicities and metallicity gradients

M. Steinmetz[1,2] and E. Müller[1]

[1] Max–Planck–Institut für Astrophysik, Karl–Schwarzschild–Straße 1, 85740 Garching b. München, FRG
[2] Institut für Theoretische Physik und Sternwarte der Universität Kiel, Olshausenstraße 40, 24098 Kiel, FRG

**Abstract.** We present first results concerning the metallicities and stellar populations of galaxies formed in a cosmologically motivated simulation. The calculations include dark matter, gas dynamics, radiation processes, star formation, supernovae feedback, and metal enrichment. A rotating, overdense sphere with a mass of $8\,10^{11}\,M_\odot$ serves as initial model. Converging and Jeans unstable regions are allowed to form stars, which get their metallicity from the gas they are formed from. Via supernovae, metal enriched gas is given back to the interstellar medium. The forming galaxy shows the main properties of spiral galaxies: A metal rich bulge, a metal poor stellar halo and a disk of nearly solar composition. Halo and bulge consist predominantly of old stars ($> 10.5\,\text{Gyrs}$). The disk has a metallicity gradient of $d(\log Z)/dr = -0.05\,\text{kpc}^{-1}$, whereas the halo shows none. The models also exhibit a correlation between the metallicity of Pop II stars and the power of small scale fluctuations. The stars of the bulge form from gas which is initially located in the largest maxima of the primordial density fluctuations, whereas the halo stars originate from gas accumulated in less pronounced maxima.

## 1. Introduction

During the last two decades several models have been proposed to explain the morphological differences between ellipticals and spirals. However, most of those models ignored the influence of the underlying cosmogony for structure formation. This drawback was the focus of the pioneering work of Katz (1992), who simulated the formation of spirals in a hierarchical cosmogony including gas dynamics and star formation. Though he has included only the tidal interaction with the environment and has

*Send offprint requests to*: M. Steinmetz

completely neglected mass exchange, in his simulations objects formed which have the main properties of observed spiral galaxies: a disk, a stellar spheroid and a dark matter halo. Most recently Navarro & White (1993, 1994) have done similar simulations starting from fully consistent CDM initial conditions. However, in their approach a larger (factor of 2) computational volume must be considered. Using the same number of particles this implies a factor of 2 smaller spatial resolution. We have done simulations similar to that of Katz (1992), but have included the metal enrichment due to supernovae. In contrast to the calculation of Katz (1992) and of Navarro & White (1993), we did not use a simple equilibrium cooling function. Instead we followed in detail the non-equilibrium ionisation of H and He as well as the radiation background in a way similar to that described in Cen (1992; see also Steinmetz & Müller 1994a). The dynamical evolution of the system as well as the structure and kinematics of the final object and their dependence on the model parameters are subject of a set of forthcoming publications (Steinmetz & Müller 1994a, b). In this letter we focus our attention to the chemical evolution and the stellar population of a galaxy formed in such simulations.

## 2. Methods

We follow the evolution of a system of gas, dark matter and stars. The gas component is treated by smoothed particle hydrodynamics (SPH; Lucy 1977) which is described and tested in detail in Steinmetz & Müller (1993). The collisionless system of stars and dark matter is followed by a N–body code. Self gravity is included by a tree code. The resulting N–body–tree–SPH–code is similar to that of Hernquist & Katz (1989) and of Navarro & White (1993) and will be described in detail in a forthcoming publication.

The star formation algorithm is similar to that used by Katz (1992) and Navarro & White (1993), i.e., converging, Jeans unstable and rapidly cooling regions are allowed to form stars. About one third of the mass of an unstable SPH (pseudo) gas particle forms a new collisionless star particle, while the other two third of the gas mass are assumed to be heated by super-

Steinmetz & Müller 1994a). In contrast to the usual implementation, we do not take all the mass of the new star particle from only one gas particle, but from all gas particles overlapping the new star particle according to their SPH–contribution to the local gas density. After a star particle is formed, we assume that the mass distribution of the about $10^7$ single stars which are represented by a star particle is given by a Miller–Scalo like initial mass function (IMF, Miller & Scalo 1979). For simplicity we assume: i) the IMF as well as the upper and lower mass limits is be constant in time. ii) Stars more massive than $8\,M_\odot$ end as type II supernovae, which are probably the dominant source of metals and energy for the chemical evolution of a galaxy. iii) With exception of a $1.4 M_\odot$ remnant all of their mass is assumed to be given back to the interstellar medium.

The metal enrichment due to type II supernovae is modelled as follows: Every gas and star particle is labeled by a metallicity $Z$, which initially is zero. Every supernovae with a progenitor mass of $m$ solar masses synthesizes $(0.357\,m - 2.2)$ solar masses of heavy elements (Maeder 1987). Integrated over the high mass end of the IMF, this imlies that the ejecta of a supernova are enriched by 15% in mass with metals. The enriched material is spread across the neighbouring gas particles according to the SPH–formalism. Vice versa, the metallicity of the next generation of star particles is the SPH–weighted sum of the metallicities of the gas particles the form from. As a consequence, the gas is successively enriched with metals. Moreover, stars which form later in the evolution are in general more metal rich than those formed earlier on.

The lifetime of stars more massive than $8\,M_\odot$ is less than 25 million years. Furthermore, the time interval between the onset of the gravitational collapse and the formation of a star can also be of the order of a few million years. The timescales of both processes is shorter than the time resolution of our star formation scheme ($\Delta t = 10^7$ yrs). We, therefore, assume, that the interstellar medium is enriched by mass, energy and metals from a newly born star particle at a constant rate over a time interval of $3\,10^7$ years, i.e., 3 star formation timesteps $\Delta t$. This ensures a smooth transition from non active to active star formation and vice verca. After $3 \cdot 10^7$ yrs, there is no further mass, metal and energy input to the interstellar gas. As in Katz (1992) the energy output is added to the thermal energy of the gas. In summary, a newly born star particle of mass $m$ returns by supernovae 12% of its mass to the interstallar medium. 15% of this mass is in form of metals.

## 3. Results

In the following we describe the results of our calculations concerning the metal enrichment. The calculations start at a redshift of $z = 25$. An initially rigidly rotating sphere with a mass of $8\,10^{11}\,M_\odot$, a radius of 50 kpc (1.3 Mpc in comoving coordinates) and a spin parameter of $\lambda = 0.08$ serves as initial model. The spin parameter is typical for a dark halo in a hierarchical clustering scenario. Th e choice of mass and radius corresponds to a $3\sigma$ overdensity in a biased CDM–scenario ($b = 2$) on a mass ($H_0 = 50$ km/sec/Mpc, $\Omega_0 = 1$). On the matter in our computational sphere we have imposed small scale density fluctuations according to a CDM–spectrum. The calculations stop at $z = 1$. At that time, the formation of a disk galaxy is completed and a quasi stationary state is reached. The galaxy consists of more than 25000 star particles making it prohibitively expensive to continue the calculation up to a redshift of $z = 0$. For a detailed discussion of the initial conditions, of the dynamical evolution and of the kinematics and structure of the final object we refer to forthcoming publications (Steinmetz & Müller 1994a,b) and to Katz (1992).

In Fig. 1 the metallicity $Z$ of the stars is shown as a function of their age (with respect to $z = 0$). We have descriminated between bulge (left), disk (middle) and halo stars (right). A star particle is called a bulge star, if its radial distance to the center $r$ is less than 1 kpc. Disk stars are those particles, for which the distance to the rotational axis $R$ is larger than 3 kpc and which height $|z|$ above the disk is less than 1 kpc. Halo particles have $|z| > 3$ kpc. We should note here, that these definitions of bulge, disk and halo are in good agreement with other indicators derived from the kinematics and the structure of the galaxy (see Steinmetz & Müller 1994b).

On the left plot of Fig. 1 one can see, that the metallicity of the early bulge stars increases very rapidly. Within 500 million years, a metallicity of about $2\,Z_\odot$ is reached. Nearly all bulge stars younger than 11 Gyrs have more than solar metallicity. These relatively young stars should, however, be rather classified as the inner disk than as bulge stars.

The disk stars in the middle of Fig. 1 are mostly younger than 11 Gyrs. Their metallicity increases more slowly over a time interval of about 2 Gyrs. The scatter in $Z$ is much larger than in the bulge, ranging from $0.2\,Z_\odot$ to $2\,Z_\odot$. However, about 13% of the disk stars have metallicities less than $0.25\,Z_\odot$, in contrast to the observational limit of 2% in the solar neighbourhood (for a review, see e.g., Audouze & Tinsley 1976), indicating a G–dwarf–problem. One should note, however, that in the 8.5 Gyrs from $z = 1$ to $z = 0$ further stars will be formed out of the metal enriched gas. If we extrapolate to $z = 0$ assuming a global star formation rate of about $1\,M_\odot$/yr, which is roughly the global star formation rate at the end of the simulation, the mass fraction of metal poor stars in the disk will decrease to less than 9%. In addition, the mass fraction of stars with metallicities less than $0.1\,Z_\odot$ is already less than 2%. Therefore, within a factor of 2, which is certainly consistent with the errors of our poor metal enrichment scheme, we can avoid a G–dwarf–problem.

The halo stars in the right plot of Fig. 1 are mainly formed during the first 2 Gyrs. Their metallicity ranges between 0 and $0.5\,Z_\odot$. However 10% of the halo stars have $Z < 10^{-3}\,Z_\odot$. The few halo stars younger than 10.5 Gyrs disappear if we take a more restrictive halo criterion, like e.g., $|z| > 5$ kpc. A further increase of the height criterion restriction leaves the age distribution $10.5\,\text{Gyrs} < t_{\text{halo}} < 13\,\text{Gyrs}$ unchanged, only the number of halo stars with the largest metallicities $Z > 0.1\,Z_\odot$ is reduced.

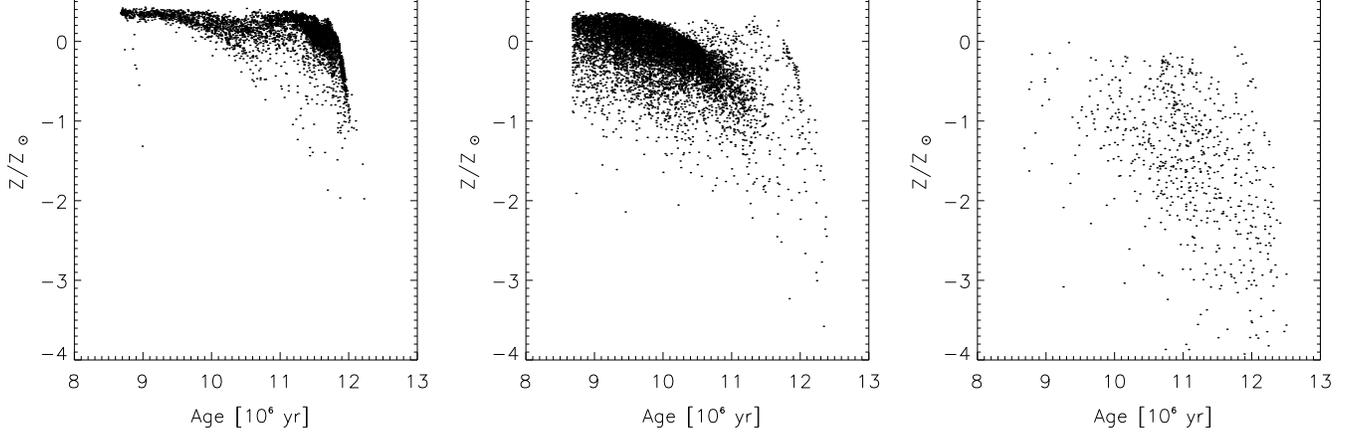

**Fig. 1.** Metallicity of the stars as a function of their age at a redshift of $z = 1$. Left: bulge stars, middle: disk stars, right: halo stars.

In Fig. 2 the metallicity is shown as a function of radius at $z = 1$, binning the metallicity of the stars into cylinders concentric to the rotational axis. The dashed line shows the metallicity of the disk stars, the full line is a best fit corresponding to $d(\log Z)/dR = -0.05\,\mathrm{kpc}^{-1}$, in good agreement with observations (Maciel 1993). The dash dotted line shows the metallicity of the halo component. Almost no gradient is visible up to distances of 25 kpc, which is in agreement with observations, too (Zinn 1985). The metallicity of the bulge component (not shown in Fig. 2) seems to obey a power law $d(\log Z)/d(\log r) \approx -0.1$. This result, however, might be affected by the limited numerical resolution (1kpc).

In Fig. 3 we have examined whether or not there exists a correlation between the metallicity of the stars and the maxima of the primordial density fluctuations. Not shown in the plot are all stars younger than 10.5 Gyrs, whose metallicity seem to be uncorrelated with the primordial density fluctuations. This can well be understood, because old stars ($t > 10.5\,\mathrm{Gyrs}$) are mainly formed in local density maxima, whereas younger stars are formed during the collapse of the entire system itself. In the middle plot of Fig. 3 one can see that the bulge stars are preferentially formed in the main maximum of the primordial fluctuation field. On the other hand gas particles, which end up in old metal poor halo stars, are more diffusively distributed (right frame of Fig.3). Obviously, there exists a strong correlation between the metallicity of the old stars and the strength of the primordial density fluctuations. A possible explanation of the physical mechanism for this correlation may be feedback limiting star formation in weaker potential wells. Consider a weak density maximum in the gas forms a star. Then the heat input from the first supernovae will prevent further star formation, i.e., only a few generations of stars are formed, whereas in the more prominent maxima more generations of stars can be formed before supernova heating stops the infall of the gas. Consequently, self enrichment of the gas should play a dominant role. More detailed investigations are necessary, however, to put this scenario of this feedback limiting star formation on a firm basis.

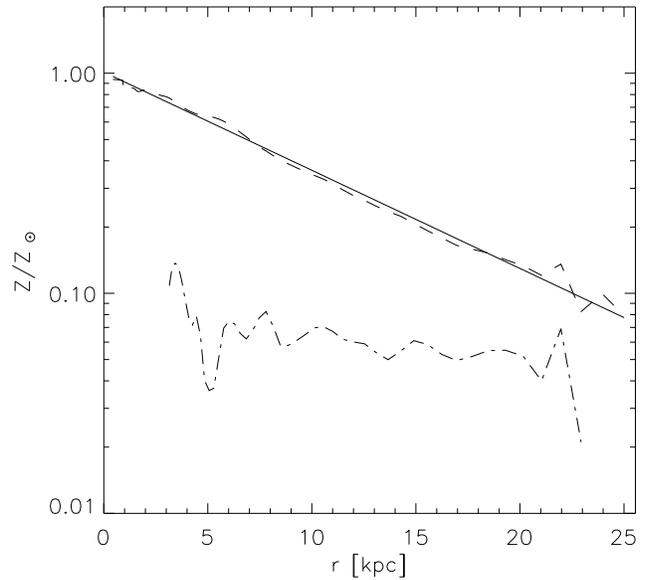

**Fig. 2.** Metallicity of stars as a function of radius. The dashed line shows the metallicity of disk stars and the dash–dotted that of halo stars ($|z| > 3\,\mathrm{kpc}$, $t_* > 2\,\mathrm{Gyrs}$). The solid line represents a least square fit to the metallicity of disk stars the gradient being $d(\log Z)/dR = -0.05\,\mathrm{kpc}^{-1}$. Note that the halo stars shows almost no metallicity gradient.

## 4. Conclusion

We have investigated the metal enrichment of galaxies formed in a environment modelled according to the hierarchical clustering scenario. The model provides information on the age, the distribution and metal content of the forming stars. From our results it is easily possible to distinguish Pop I and Pop II stars in the simulated galaxies. As observed the simulated galaxy shows a metal rich bulge consisting of old stars, a disk of about solar metallicity and a metal poor halo of old stars. The metallicity of the disk stars has a gradient of $d(\log Z)/dR = -0.05\,\mathrm{kpc}^{-1}$, the old halo component posseses almost no gradient. One possible explanation of the differences between the metallicity of the old Pop II stars in the bulge and the stellar halo is due to the combined effects of small scale primordial density fluctuations and self enrichment. The fact that the galactic components (bulge, disk and halo) identifable by their chemical and population properties are rather similar to those components obtainable by an analysis of the structure and the kinematics of the galaxy (Katz 1992, Steinmetz & Müller 1994a) strongly points towards the picture of a hierarchical galaxy formation.

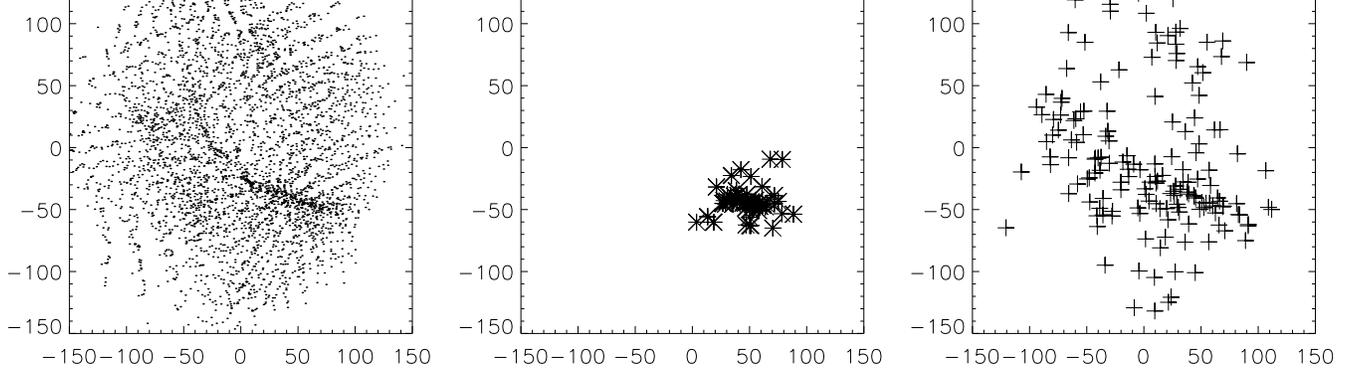

**Fig. 3.** Snapshot of gas particle distribution in a forming spiral galaxy at a redshift of $z = 6$. Left frame: Projected distribution of all gas particles in the x–y–plane. Middle frame: Position of gas particles which will form old and metal rich (bulge) stars. Right frame: Position of gas particles which will form old and metal poor (halo) stars.

*Acknowledgements.* We would like to thank M. Bartelmann, A. Burkert, G. Hensler and C. Theis for many fruitful discussions. We also acknowledge useful comments and suggestions by S. White. This work is partially supported by the *Deutsche Forschungsgemeinschaft*. All calculations were performed on the Cray YMP 4/64 at the Rechenzentrum Garching.